\documentstyle [aasms4]{article}
\def\kms { km\,s$^{-1}$ }
\def\hipass {H{\sc i}PASS }

\lefthead{Banks et al.}
\righthead{New Galaxies in the Cen\,A Group}
 
\begin{document}

\title{NEW GALAXIES DISCOVERED IN THE FIRST BLIND H{\sc i} 
SURVEY OF THE CENTAURUS A GROUP}

\author{G. D.  Banks\altaffilmark{1} M. J. Disney\altaffilmark{1} P. M. Knezek\altaffilmark{2}$^\ddag$ H. Jerjen\altaffilmark{3} D. G. Barnes\altaffilmark{4} R. Bhatal\altaffilmark{5} W. J. G. de Blok\altaffilmark{6} P.J. Boyce\altaffilmark{1} R. D. Ekers\altaffilmark{4} K. C. Freeman\altaffilmark{3} B. K. Gibson\altaffilmark{7} P. A.  Henning\altaffilmark{8} V. Kilborn\altaffilmark{6} B. Koribalski\altaffilmark{4} R. C.  Kraan-Korteweg\altaffilmark{9} D. F. Malin\altaffilmark{10} R.F. Minchin\altaffilmark{1}  J. R. Mould\altaffilmark{3} T. Oosterloo\altaffilmark{11} R. M. Price\altaffilmark{4} M. E. Putman\altaffilmark{3} S. D. Ryder\altaffilmark{12} E. M. Sadler\altaffilmark{13} L. Staveley-Smith\altaffilmark{4} I. Stewart\altaffilmark{4} F. Stootman\altaffilmark{5} R. A. Vaile\altaffilmark{5}$^\dag$ R. L. Webster\altaffilmark{6} A. E. Wright\altaffilmark{4}}

\altaffiltext{1}{University of Wales, Cardiff, Department of Physics \& Astronomy, P.O. Box 913, Cardiff CF2 3YB, Wales, United Kingdom}

\altaffiltext{2}{The Johns Hopkins University, Department of Physics \& Astronomy, 34$^{\rm th}$ \& North Charles Streets, Baltimore, MD 21218, USA}

\altaffiltext{3}{Research School of Astronomy and Astrophysics, Australian National University, Weston Creek P.O., Weston, ACT 2611, Australia.}

\altaffiltext{4}{Australia Telescope National Facility, CSIRO, P.O. Box 76, Epping, NSW 2121, Australia}

\altaffiltext{5}{University of Western Sydney Macarthur, Department of Physics, P.O. Box 555, Campbelltown, NSW 2560, Australia}

\altaffiltext{6}{University of Melbourne, School of Physics, Parkville, Victoria 3052, Australia}

\altaffiltext{7}{Center for Astrophysics and Space Astronomy, University of Colorado, Boulder CO, 80309-0389, USA.}

\altaffiltext{8}{University of New Mexico, Department of Physics \& Astronomy, 800 Yale Blvd. NE, Albuquerque, NM 87131, USA}

\altaffiltext{9}{Departamento de Astronomia, Universidad de Guanajuato, Apartado Postal 144, Guanajuato, GTO 36000, Mexico}

\altaffiltext{10}{Anglo-Australian Observatory, P.O. Box 296, Epping, NSW 2121, Australia}

\altaffiltext{11}{Istituto di Fisica Cosmica, via Bassini 15, I-20133, Milano, Italy }

\altaffiltext{12}{Joint Astronomy Center, 660 North Aohoku Place, Hilo, H{\sc i} 96720, USA}

\altaffiltext{13}{University of Sydney, Astrophysics Department, School of Physics, A28, Sydney, NSW 2006, Australia}

\noindent $^\ddag$Guest observer, Cerro Tololo Inter-American Observatory, which is operated by AURA, Inc.\ under contract to the National Science Foundation

\noindent $^\dag$ deceased

\begin{abstract}

We have commenced a 21-cm survey of the entire southern sky ($\delta<0^\circ$,
$-1200$\kms $<v_{\odot}<12,700$\kms) which is ``blind'', i.e. unbiased by previous 
optical information.  In the present paper we report on the results of a pilot 
project which is based on data from this all-sky survey. The project was carried 
out on an area of 600 square degrees centred on the nearby Centaurus A (Cen\,A) 
group of galaxies at a mean velocity of $v_{\odot}\sim$ 500 \kms. This was 
recently the subject of a separate and thorough optical survey.  

We found 10 new group members to add to 21 
    galaxies already known in the 
Cen\,A group: five of these are previously uncatalogued galaxies,    
  while 
five were previously catalogued but not known to be associated with the group. 
Most of the new members have H{\sc i} masses close to our survey limit 
of 10$^{7}$ M$_{\odot}$ at the assumed group distance of 3.5 Mpc. 
The  new detection with the largest H{\sc i} mass is ESO174-G?001 with 
$M_{H{\sc i}}=2.1\times 10^{8}$ M$_{\odot}$. 
 Prior to  our survey this galaxy 
was an uncertain optical identification because of high galactic extinction. 
  We found  optical counterparts  for 
 all  the H{\sc i} detections, most of them 
intrinsically very faint (M$_{B}<-13.0$) low-surface brightness dwarf 
galaxies with  H{\sc i} profile line-widths suggestive of dynamics dominated 
by dark matter. The new group members add approximately 6\% to the H{\sc i} 
mass of the group, and  4\% to its light. The H{\sc i} mass function,  derived
from all   the  known group galaxies  
   in   the  interval 
  $10^{7}$ M$_{\odot}< M_{\rm H{\sc i}}< 10^{9}$ M$_{\odot}$, 
  has  a faint-end slope of 1.30 $\pm$ 0.15, allowing us to rule out a slope
of 1.7 at 95\% confidence.  Even if the number in the lowest mass bin is 
increased by 50\% the slope only increases to 1.45 $\pm$ 0.15.

\end{abstract}


\section{Introduction}
Recent developments in technology make it possible to perform for 
the first time large-area sensitive blind surveys at 21-cm to search
for extragalactic H{\sc i}. This opens many scientific opportunities. Most 
importantly, the selection effects through the 21-cm window will be quite different 
from the well known optical selection effects which may presently bias 
our understanding of galaxy populations (e.g. Impey \& Bothun 1997, 
Disney 1998). For instance, low surface brightness (LSB) galaxies are 
dimmer than the sky and nearby gas-rich dwarf galaxies show a morphology
confusingly similar to background giant galaxies which make both galaxy
types very difficult to  identify optically.  But they may well become
``visible'' in a 21-cm survey if  they    have  a significant amount 
of H{\sc i}. LSB galaxies appear to be common - with approximately equal numbers of 
galaxies, of a given size, in each one-magnitude bin of blue central 
surface brightness $\mu_{B,0}$     
measured between $21.5<\mu_{B,0}<26.5$ (Turner 
et al.~1993; McGaugh 1996). Considering the relative narrowness of 
the bins at fainter magnitudes this is a remarkable result. The 
cosmological significance of LSB galaxies remains controversial
because of uncertain  optical selection effects at faint luminosities 
(see Impey \& Bothun 1997, Disney 1998 
   and references therein). Likewise there is 
serious disagreement about the relative number of faint dwarf 
galaxies. Magnitude-limited surveys tend to find a flat faint-end slope 
(Schechter, 1976, 
   parameter $\alpha \sim 1$) to the optical luminosity function 
(e.g. Loveday et al.~1992, Lin et al.~1996). This would suggest that the 
total light contribution from dwarfs in the universe is insignificant, 
whereas surveys within clusters of galaxies (e.g.  
   de Propris et al. 1995;  Driver \& Phillipps 1996;  
Trentham 1997;  Phillipps, 
 Parker, Schwartzenberg \& Jones 1998) find steep faint end slopes with $\alpha>2$, suggesting a  
dominant  role for dwarfs in the determination of the light content of 
the universe. Dwarf galaxies may also contain appreciable amounts of 
dark matter (e.g. Freeman 1997), implying that they may add even more to the 
overall mass in the universe than to the light. Finally, any extragalactic survey 
in which every object is automatically tagged with its velocity-distance 
offers rare opportunities for statistical studies. 

Blind 21-cm surveys have a long history but their ambitions 
have been severely curtailed by technological limitations. 
Single dish radio telescopes large enough to reach high sensitivity 
have had beam areas too small to survey large areas of sky, while 
interferometric surveys are still confined to relatively narrow 
velocity windows by the computer power required to carry out 
the necessary cross-correlations between every interferometer 
pair. Nevertheless,  in recent times 
  blind surveys have been carried out in 
the galaxy clusters Hydra by McMahon  et al.~(1993) and Hercules
by Dickey (1997); in voids by Krumm \& Bosch (1984) and 
Weinberg et al. (1991);  behind the Galactic plane by Kerr \& 
Henning (1987); and in the field by Henning (1995). 
 Latterly, narrow zenith-strip scans have been done at Arecibo  
  (see later)  by 
Spitzak \& Schneider (1998) and Zwaan et al.~(1997).

The most recent 21-cm survey is our  H{\sc i} Parkes All-Sky Survey 
(\hipass) which started early in 1997 and is expected to be 
complete within $\sim$3 years. This project takes advantage of the 
new multibeam instrument at the 64m radio telescope at 
Parkes (Australia) equipped with 13 separate beams (26 receivers) and 
enough correlator power in each beam to survey 1024 velocity channels 
simultaneously. The scientific aims of the project include a  
 study  of  the Galactic H{\sc i} and  of the galaxy content of the local 
universe. For this purpose     we scan  
  the complete southern sky in the 
velocity range $-1200$\kms $<v_{\odot}<12,700$\kms. 

For the purpose of testing the performance of \hipass  the present 
limited study was carried out using  the 
   first set of \hipass   data completed
to full sensitivity. The selected area covers a substantial fraction of the nearby 
Centaurus A group of galaxies. The main reasons for this choice are twofold:  
   (1) 
the Cen\,A group is among the closest galaxy groups to the Local Group
at $\sim 3.5$\,Mpc (Hui et al.~1993; Sandage et al.~1994) and 
contains the richest population of   different morphological galaxy types (de 
Vaucouleurs 1979), including the peculiar giant elliptical NGC 5128, the 
large disk galaxies M 83 and NGC 4945, the S0 galaxy NGC 5102, and the active 
star forming galaxy NGC 5253. All main group members appear disturbed in 
their morphology. It has been suggested (Graham 1979; van Gorkom et al.~1990) 
that the recent accretion of a population of gas-rich dwarfs could be the 
reason.  The group lies in the Supergalactic Plane with its southern 
extension running down to low galactic latitudes making optical studies 
challenging because of high extinction. (2) The group has recently been the 
 subject of a careful optical survey (C\^ot\'e et al.~1997)  enabling 
 comparison between contemporary optical 
 searches for gas-rich type (irregular) 
 galaxies and   our ``blind'' H{\sc i} 
  survey.   

 Most 
   earlier  H{\sc i} studies of groups of galaxies have been done based on  
  previous  
optical identifications of candidate members.  In the M~81 group,  
van Driel et al.\ (1998) and Huchtmeier \& Skillman (1998) have independently 
done 21~cm searches for optically selected dwarf candidates.  Neither group 
detected any new members. 
 Huchtmeier, Karachentsev \& Karachentseva (1997) followed up on  
potential dwarfs found on the Second Palomar Sky Survey (POSS II), and found 
1 new candidate of the IC~342/Maffei group, and three new dwarf members each 
in the NGC~6946 and NGC~672/IC~1727 groups.   
Maia \&  Willmer (1998) have studied the neutral 
hydrogen content of a number of ``loose groups'', but they also used  
pointed observations of optical candidates.

   Several H{\sc i} searches of  
compact groups of galaxies have been done (Oosterloo \& Iovino 1997; Williams 
\& van Gorkom 1997; Williams, McMahon \& 
  van Gorkom 1991), however these groups are usually 
far enough away that both the spatial resolution and the sensitivity preclude 
detecting dwarf galaxies.   However, a 
   few nearby groups have been surveyed 
 blind at 21~cm, and they include 
M~81, CVnI and  NGC~1023, in which Lo \&  Sargent (1979) found 4 
 previously uncatalogued members, and a part of  the 
 Sculptor   group, studied by 
 Haynes \& Roberts (1979) who found  uncatalogued H{\sc i} 
objects thought to be  group members.

The present paper is organised as follows: \S 2 describes the 21-cm survey;  
the observations with results are given in \S 3 and \S 4. The optical
follow-up on the newly detected galaxies and their optical properties are 
presented in \S 5 and \S 6. In \S 7 we derive the H{\sc i} Mass Function
for the Cen\,A group. Finally, we discuss detection limits and future projects 
in \S 8.   

\section{Survey}

   The Cen\,A group  covers approximately 1000 square degrees of sky down to and perhaps 
 through  the galactic plane. Since it is also projected  against 
 the Supergalactic  plane, the area is rich in galaxies and therefore 
 well suited to test the performance of \hipass.  

  The comparison optical search for dwarf irregulars in the area by 
C\^ot\'e et 
al.~(1997; hereafter C97)  covered 900 square degrees 
between   
  $12^{h}30^m< \alpha <15^{h}00^m$ 
   and $-50^{\circ}< \delta <-20^{\circ}$ whereas we have  surveyed a 
 smaller  area of 600 square degrees, chosen because it fits 
 into the \hipass  survey grid. C97 found 18 H{\sc i} rich dIrr's 
(2 new)  of which 14 were in our search-area.  Adding 
 AM 1321-304  found by  Matthews \& Gallagher (1996) and the six well known 
 bright group members NGC 5128, M 83, NGC 4945, NGC 5102, 
 NGC 5253 and ESO 270-G017  means there were in all 21 
 suspected group members in our 600 square degree 
 search area prior to our survey, with another 7 
 outside (see Figure 1). Our search was  confined to 
 the velocity range 200 $< v_{\odot} <$ 1000 \kms, cut off at 
 the lower end because of the possibility of 
 confusion with High Velocity Clouds which are numerous in 
 the Cen\,A group direction (Wakker, 1991). One 
 group member - CEN 5 ($v_{\odot}$=160 \kms) 
 with a velocity lower than 200 \kms has been previously claimed  (C97) and there remains 
 the possibility  that other group members in our data  remain excluded by the 200 \kms 
 restriction.

  If we exclude CEN 5 from the statistics  then  there were in our search 
 area, prior to this survey, 18 suspected group members with H{\sc i} 
detections and two with H$\alpha$ (NGC 5206 and 
 ESO 272-G025).

\section{ Observations and Data Reduction}
The 21-cm observations for this pilot study were carried out in February and 
March 1997 as part of the  \hipass survey which is ongoing  at the 64m 
single dish Parkes Radio Telescope. This data set comprises the first 
completed area observed to full sensitivity with \hipass. The instrument is a 
multibeam system consisting of 13 beams, each with 2 orthogonal polarisations. The 
scanning rate across the sky is 1 degree/minute at constant right ascension. 
The data from the 26 correlators are stored every 5 seconds. Each correlator has 
1024 channels tuned to a velocity range between $-1200$ \kms to $12,700$ \kms, 
 corresponding  to a channel separation of $13.2$\kms. The 
   system temperature, $T_{\rm sys}$, is $\sim 21$ K and the 
  scanning pattern is 
organised so 
   that every point in the sky passes through several beams during a scan, 
and through several different scans separated by days or weeks of time depending on 
the observing schedule. This allows for the rejection of time-dependent interference. 
The total integration time per point on the sky is 450 seconds. The 5$\sigma$ sensitivity of the 
\hipass survey is measured at 70 mJy per channel per beam. For more details 
about the multibeam system we refer to Staveley-Smith et al.~(1996).  
The observed spectra were smoothed on-line by applying a 25\% Tukey filter, to  
reduce ``ringing'' caused by the strong Galactic signal entering through spectral 
side lobes. This process accounts for a loss in velocity resolution of 37\%. 
Consequently the actual velocity resolution is $18.2$\kms, or $35$\kms after 
Hanning smoothing. A description of the on-line data reduction at the telescope 
which involves bandpass correction can be found in Barnes et al.~(1998). 
 Three-dimensional data cubes ($\alpha$, $\delta$, $v_{\odot}$)
were produced by putting individual spectra onto a pixel grid of 
$4^{\prime} \times 4^{\prime}$. Due to the lack of an automatic source finder 
at this early stage of the \hipass survey, these cubes were visually searched 
for H{\sc i} sources by moving through the velocity range $0<v_{\odot}<1000$ \kms,   
which covers the velocity distribution of the Cen\,A group,  and corresponds 
   to 
1000\kms/35\kms=28 independent velocity resolution elements at a given 
spatial pixel on the sky. Given that there are 10,000  
  such  pixels in this survey, we searched 
through 280,000 independent ``cubicles''  (i.e. 3-dimensional pixels). 

  The 1$\sigma$ sensitivity of 
the \hipass survey  was   measured at  
   14 mJy/channel. Assuming  noise  normally distributed  about 
  zero, we expect 
 statistically to find $\sim$ 0.1  spurious positive 5-sigma detections in 
the data. However, 27 objects with velocities $v_{\odot}>200$\kms    
were found at this detection threshold (plus CEN 5). 
 Among others we recovered  main Cen\,A group 
members such as NGC 4945, NGC 5102, M 83, NGC 5237, and NGC 5253. NGC 5128 remained undetected 
because  the H{\sc i} emission was swamped by  
  strong continuum emission from that galaxy.

Many more H{\sc i} sources have been identified with velocities $v_\odot < 200$\kms.
These are most likely not galaxies but galactic high velocity clouds,  as 
there is a clear overlap in velocity space between nearby galaxies and the system of 
HVCs of our Galaxy towards  the Cen\,A group direction (Wakker 1991). 
Because of the number dominance of high velocity clouds,  and because only one Cen\,A 
group galaxy is known below $200$\kms to date (CEN 5, see Table 1),  
   we 
restricted our study and optical follow-up program (see \S 5) 
   to   detections with 
$v_\odot>200$ \kms.

\section{H{\sc i} properties}

We determined the H{\sc i} fluxes of our  detections by making zero-th 
moment maps. A MIRIAD (Sault et al.~1995) gauss fitting routine yielded an estimate 
of the object's coordinates  $(\alpha,\delta)$ (cols.~[2] and [3] in Table 
 1).   
Positional accuracy should equal the half-power beamwidth ($14^{\prime}.8$) divided by 
the signal-to-noise ratio, which corresponds to $< 3^{\prime}$ (3.2 kpc at the 
distance of Cen\,A) for all sources with a 5$\sigma$ detection or better. A 
spectrum was extracted from the data cube at the sky position from which the 
heliocentric radial velocity, $v_{\odot}$ (col.~[4]), was measured at the mid-point 
at 50\% of the peak of emission, and the total flux was determined (col.~[5]). 
Furthermore, we measured the velocity widths $\Delta V_{50}$ (cols.~[6]) and 
$\Delta V_{20}$ (cols.~[7]) at 50\% and 20\% of the peak, respectively. The 
H{\sc i} mass (col.~[8]) was computed by employing the standard formula: 

$$
\frac {M_{\rm H{\sc i}}}{M_{\odot}} = 2.356 \times 10^{5} (D[{\rm Mpc}])^{2} 
\int{S_{\nu}dV} 
$$

\noindent
where D is 3.5 Mpc, our assumed distance to the 
   Cen\,A   group.

The examination of the Hanning smoothed 
 spectra (Figure 2) reveal that none of the detections are 
marginal. All sources are found to have signals above 3.9 Jy\kms. 
  The coordinates of the H{\sc i} detections were cross-correlated with 
the NASA Extragalactic Database and compared with UK Schmidt films to find optical 
counterparts. Each source could be associated unambiguously with  a galaxy 
(col.~[1] of Table 1 
 ) which is located within the positional 
   uncertainty 
of 3$^{\prime}$ radius. In some cases more accurate position informations were  
  acquired with the Compact Array of the Australia Telescope (see Staveley-Smith et al.~1999)  where   the identifications were 
 always confirmed.  

Among the 28 detection we found ten new Cen\,A group members. 
  Five of these  galaxies were 
previously catalogued but not known to be associated with  the group, 
 while the  other five
were previously uncatalogued galaxies and are identified  according to the \hipass 
survey nomenclature (HIPASS 1321-31 has recently been detected as 
 131820.5-311605 by  Karachentseva \& Karachentsev  1998). 

The most gas-rich new group member is ESO 174-G?001 with a H{\sc i} mass of 
$M_{H{\sc i}}=2.1\times 10^{8}$ M$_{\odot}$,  
   which is larger than the amount of 
H{\sc i} observed for instance in NGC5253 or NGC5408. It is a low surface brightness 
galaxy close to the Galactic plane ($b=8^{\circ}$) which may explain why its identification as a 
galaxy was previously uncertain (Lauberts 1982). CEN 5 was previously optically 
identified and classified as a group member candidate by C97, and so 
 we include the spectrum for completeness, although it is excluded from the 
 statistics by its low velocity ($<$ 200 \kms).   
  Looking at the spectrum (Fig. 2), we see two peaks. However, 
 the   higher velocity   signal is from the   nearby  
   galaxy  NGC 4945,   the complete profile of 
   which  can   be seen    in Figure 2.

For completeness reasons we also list in Table 1
  the H{\sc i} limits 
NGC 5206, ESO 272-G025, and NGC 5128 which remained undetected in this study: NGC 5206
is an early-type galaxy and ESO 272-G025 was classified as peculiar (Lauberts, 1982).   These  two galaxies  were also not detected in  H{\sc i} emission 
  by C97, to a more sensitive detection limit.  
Since the Multibeam failed to detect the powerful radio galaxy NGC 5128, 
 the listed H{\sc i} mass for this galaxy 
is taken   from de Vaucouleurs et al. (1991),  derived from absorbing gas. 

In Figure 3 and Figure 4 we compared the total fluxes and velocity-widths 
($\Delta$V$_{20}$) with the corresponding data measured by C97. 
   In   general we found
good agreement,  showing H{\sc i}PASS is achieving 
 the expected sensitivity,   although for three galaxies we measured   larger integrated fluxes, suggesting that there may be a 21-cm signal 
 outside C97's chosen search areas.

\section{Optical properties}

We attempted to get CCD images in the optical photometric bands $B$ and $R$ for 
those galaxies without reliable photometry in the literature. The optical data
was acquired as part of an extensive optical follow-up program of \hipass
galaxies. This program employs three telescopes: (i) the 40-inch telescope at 
the Siding Spring Observatory, (ii) the 1m Swope Telescope at Las Campanas Observatory,
and (iii) the Curtis Schmidt telescope at CTIO. 
 We used $2048^{2}$ CCDs with a $\sim$ 24$^{\prime}$ field of 
view and  a pixel size of $0.7^{\prime\prime}$, except for the 
 Siding Spring  observations, where 
 a 800$^{2}$  CCD was used with a $\sim$ 8$^{\prime}$ field of 
view and the same pixel size.  For each galaxy a series
   of 6 exposures of 300 
seconds each were taken.   The optical data were  
 de-biased and flattened 
 with the standard STARLINK CCDPACK (Draper 1993) 
 reduction techniques,   then calibrated using 
  Landolt (1992) standards.   
 The seeing was typically 
  1$^{\prime\prime}$.  
In the case of \hipass1337-39, we had to rely on the image from the 
Digitised Sky Survey, DSS\footnotemark  (a   similar   method   to that  
    employed by   Spitzak \& 
 Schneider, 1998)    to calibrate   an  image for which  we  
  had no  calibrated   CCD 
  data.  
\footnotetext{The Digitized Sky Surveys  
  were produced at the Space Telescope Science Institute under U.S.  
Government grant NAG W-2166.}

Figure 5 shows $B$-band images of the newly identified Cen\,A group galaxies 
to illustrate the variety in optical morphology. Previously catalogued 
galaxies (single *) 
are high surface brightness objects and thus likely to be visually identified as 
background galaxies. The uncatalogued galaxies  
  (**),   on the other hand, are faint, low 
surface brightness galaxies with small angular size and very hard to see.  

In Table 2 
  we list photometric parameters of the 10 new group members 
(col.[1]) drawn  from various sources - either   The Third Reference 
 Catalogue of Bright Galaxies, RC3 (de Vaucouleurs et al. 1991), the 
DSS, or from our own CCD data (if available). The coordinates (cols.~[2] 
and [3]) are the optical positions either quoted in the literature or measured by us at the 
centroid of the  optical  image. When CCD data were available, 
the galaxy image was cleaned by removing superimposed foreground stars. Afterwards, 
the total apparent magnitude $m_B$ (col.~[4]) was determined by  circular aperture 
photometry and fitting a growth curve.  
Otherwise the total apparent $B$ magnitude was 
  taken from the literature. Where 
we could fit an exponential profile $\mu(r)=\mu_0^{exp}+ 1.086(r/r_0) $ to 
our own optical data we also quote the extrapolated central surface brightness 
$\mu_{0}$ (col.~[6]) and the exponential scale length $r_0$
   (cols.~[7] and [8]). 
   In some cases the galaxy was too faint, or too obscured by  stars, 
 for a reliable profile fit to be made.

\section{Results} 
To summarise our results, a shallow   integration  H{\sc i} survey of the inner 600 square degrees of 
the Cen\,A group and the velocity range $200  <v_{\odot}<1000$\kms   
  revealed 28 H{\sc i} sources 
which all have been 
   identified as Cen\,A group galaxies. All detections have integrated 
flux larger than 3.9 Jy\kms.  
 Eighteen   objects coincide with previously known group members.  A  
  further   
 10 are 
  suggested as new group members.  Of these 10,  5 coincide with previously catalogued 
galaxies and 5 coincide with previously uncatalogued galaxies. Three main group 
members were not detected at our survey sensitivity of 10$^{7}$ M$_{\odot}$ of H{\sc i}, assuming an H{\sc i} velocity width of   
   50\kms, but one (NGC 5128) is 
 known as an  H{\sc i} source  from absorption 
 studies.    All new galaxies  detected in this study have optical 
  counterparts.

Table 3 
   lists various intrinsic parameters for the new group members. 
The optical radius $R_{opt}$ (col.~[2]) has been estimated by eye on the DSS images.  
The model-dependent total $B$ magnitudes,  m$_{T}$ (col.~[4]),  have been calculated for 
each galaxy using the parameters of the exponential profile, $\mu_{0}^{exp}$ and $r_0$ 
(cols.~[6] and [7], Table~\ref{photomone}),    using

$$
m_{T}  =  \mu_{0}^{exp} - 2.5 log(2 \pi r_0^2)
$$

  However, this  depends on an extrapolation to infinity, and so 
 the model magnitudes should be 
compared with the m$_{B}$ values in Table 2
   obtained from aperture 
photometry. The galactic extinction $A_B$ for each galaxy, given in col.(4), 
 has been obtained from the dust maps of Schlegel et al.~(1998), wherein dust is 
mapped from the energy it emits in the far infra-red as seen in the high  spatial 
resolution observations made with IRAS and DIRBE. In the case of the highly obscured 
ESO174-G?001 we have checked that estimate by measuring the colours of neighbouring background 
elliptical galaxies in our frames, and find excellent agreement ($\pm$ 0.1 magnitudes). 

Assuming a mean distance modulus of $(m-M)=27.8$ for the Cen\,A group 
  (Saha et al. 1995),  and 
correcting for the extinction, we converted the absolute magnitude M$_{B}$ (col.~[5]) 
and determined M$_{\rm H{\sc i}}$/L$_{\rm B}$ (col.~[6]).  The mass M$_{\star}$, quoted in 
col.~(7) is arrived at by simply assuming a value of the stellar mass-to-light ratio 
M$_{\star}$/L$_{\rm B}$ = 1  for such gas-rich dwarf galaxies - a result suggested 
by previous authors (e.g. Staveley-Smith et al.~1990 and de Blok, McGaugh \&  
van der Hulst, 1996). Finally, 
the ratio between indicative dynamical mass  M$_{\rm i}$ $\equiv$ R$_{opt}$($\Delta$V$_{50,corr})^{2}$/G 
(Roberts 1978) and photometric mass M$_{\star}$ is given in col.~(8).   
   $R_{opt}$   is the quoted 
optical radius   and $\Delta V_{50}$ is the 
FWHM, corrected for  inclination if the minor and major 
 axes of the galaxy could be reliably determined.

Although M$_{\rm i}$  may be a crude estimator of the true dynamical mass, it should provide 
a safe lower limit because H{\sc i} radii are usually  larger than the optical radii. For 
instance, Bosma (1978) suggests H{\sc i} disks extend out to 1.5 $\times$ the optical Holmberg 
radius, while for gas-rich galaxies, the ratio of H{\sc i} to optical diameter can be much 
greater (e.g. van Zee et al.~1995).  

The newly found group members fall in two main categories. Of the five previously 
catalogued galaxies, ESO 321-G014 (Figure 5)  and ESO 269-G058 
  (Figure 5)  are not in the C97 area: ESO 321-G014  
is obviously a dwarf irregular, while ESO 269-G058  is a highly irregular object with obvious 
H{\sc ii} regions and strong dust lanes. For the latter galaxy,  an incorrect redshift (1853 \kms) is 
published in the literature (de Vaucouleurs et al.~1991). 
 Our velocity 
for IC 4247 (418 \kms), Figure 5,  is rather  different 
  from  the value of 274 \kms   published by da Costa et 
al.~(1987), and now places it firmly in the Cen~A group. 

The five previously uncatalogued galaxies exhibit a quite different morphology. They are faint 
($\sim$ 17$^{m}$),  low surface brightness, inconspicuous objects picked up in a blind survey 
solely because they are anomalously rich in H{\sc i}. The survey wouldn't 
pick up objects as faint as m$_{B}$=17 unless they had M$_{\rm H{\sc i}}$/L$_{\rm B}$ 
ratios of 4 or more. Indeed, if we  look at their  
  ratios of   M$_{\rm H{\sc i}}$/L$_{\rm B}$   
{\bf uncorrected} for Galactic extinction we find they have values ranging  
between 4 and 10, with a median of 5.5. Higher values  are 
known in the literature (e.g. Lequeux \& Viallefond 1980; Bergvall \& 
J\"ors\"ater 1988; Brinks \& Klein 1988; van Zee, Haynes \& 
 Giovanelli, 1995)  but are very rare. If such objects have a 
stellar mass-to-light ratio M$_{\star}$/L$_{\rm B}$~=~1,  then they contain significantly 
more mass in the form of gas than of stars, and it will be interesting to investigate what 
their evolutionary status may be.

\section {The H{\sc i} Mass Function}

   There is considerable controversy  over the faint-end slope of the  galaxy 
 Luminosity Function (see \S 1) and thus added interest in the low-mass 
 end of the  H{\sc i} Mass Function (H{\sc i}MF).  
  The   H{\sc i}MF   has been recently investigated  by three 
 different  groups (Spitzak \& Schneider 1998; Zwaan et al.~1997; and Kilborn, Staveley-Smith \& Webster  1998).  The first two groups  used 
 deep pencil beam samples observed with Arecibo, whereas Kilborn 
used \hipass objects to study a flux-limited sample. Each of these  surveys  contain  between 50 
 and 100 objects  but,  because they are flux-limited, 
 they  contain  
  very few low mass objects, and are therefore  not suited to 
 determining  the low mass slope. Our survey, on the 
   other hand, being principally distance-limited  is much better suited 
 to determining  that low-mass 
 slope, though it lacks the higher masses needed to determine 
 $M^{\star}_{H{\sc i}}$ well in the usual Schechter function:

\begin{displaymath}
\Phi(M_{\rm H{\sc i}}/M^{\star}_{\rm H{\sc i}}) = 
 \Phi_{\star} 
(M_{\rm H{\sc i}}/M^{\star}_{\rm H{\sc i}})^{-\alpha} e^{-(M_{\rm H{\sc i}}/M^{\star}_{\rm H{\sc i}})}  
\end{displaymath}

   In Figure 6  we plot the H{\sc i} mass function   from the 27 detections 
  of our survey  which have accurate  H{\sc i} masses.

  In order   not to over-interpret  the data  we do not    attempt 
  to find a fit  for 
  $M^{\star}_{H{\sc i}}$, only to find   the faint-end slope $\alpha$. To 
 do this we make a weighted linear least-squares fit of 
 log $\rho$ against log $M_{H{\sc i}}$ (where $\rho$ is the number of sources per log$_{10}$ $M_{HI}$ bin) for sources with $M_{H{\sc i}} < 10^9 
 M_\odot$  (Figure 6)  and find : 

 log $\rho$  = (-0.30 $\pm$ 0.15) log $M_{H{\sc i}}$ + (3.21 $\pm$ 1.20)

 \noindent Hence the Schechter  faint-end slope is 

 $\alpha$ = 1.30 $\pm$ 0.15

 \noindent where the   formal error on the slope  (see Bevington, 1969)  
   arises from assuming  purely Poisson  statistics 
 for  the numbers in each $\rho$-bin in 
 Figure 6.    Since the most likely systematic error  is the loss 
 from our lowest mass bin of sources  which did not make the first cut, and 
 therefore  were not re-measured  more accurately, we 
 can ask how $\alpha$  would change if we added 
 50\% more sources to the 12 already in 
 that bin. In fact $\alpha$ rises 
 to only 1.45 $\pm$ 0.15.  Similarly the formal error on the slope allows 
$\alpha = 1.45$ at 1 standard deviation.
 Our data therefore does not rule out a faint-end slope of 1.5 but we
 can exclude a slope as steep as 1.7 at the 95\% level.   Table 4 
 contains a comparison  of  our Cen~A results  with those from the three other 
 H{\sc i}MF surveys mentioned above.  

\section{Discussion}

It is encouraging that the present comparatively shallow blind H{\sc i} survey  already   increases    
the number of known galaxies of the Cen\,A group by 50\%, i.e. from 20 to 30 in our search area. 
It is in particular interesting that this happened after the group was 
  so  carefully studied 
in the optical (C97).

A glance at the spectra  (Figure 2) confirms that 
none of our objects are marginal detections.   All have 
signals above 3.9 Jy\kms, whereas the formal 5$\sigma$ noise limit for 
  optimally smoothed  sources  that are   
50 \kms broad is closer to 1.5 Jy\kms. The dispersion in 
the velocity of the less significant peaks in our data map     
mimic the velocity distribution of the Cen\,A group   galaxies, 
suggesting that many of them will also turn out to be real members 
of the group.  

We find only one low surface brightness galaxy with an 
 H{\sc i} mass of more than 
   10$^{8}$ M$_{\odot}$ - ESO174-G?001  
which contains 2.8  $\times$ 10$^{8}$ M$_{\odot}$ of H{\sc i}. It is 3 
arc minutes across  and its measured  central  surface brightness 
 is  24.5 $\mu_B$, which is indeed low. However, it is the 
galaxy closest to the Galactic plane ($b=8^{\circ}$) and turns out to 
have 2.1 magnitudes of  
  foreground  blue extinction. When this is allowed for, the central 
surface brightness rises to 22.4 $\mu_B$ - which is not very low; 
M$_{\rm H{\sc i}}$/L$_{\rm B}$ drops to 0.8 and the absolute magnitude,  
M$_{B}=-15.6$.  
   ESO174-G?001  was missed by previous studies simply because 
 of the large extinction towards it.   Despite its comparatively 
large size, ESO174-G?001 is still much smaller than the instrument beam.  If there 
were a giant LSB galaxy in the group, i.e. one with a diameter larger than 
the projected beam (15 kpc), we could only have detected it, on 
signal to noise grounds, if its projected mass were more than 10$^{7}$ M$_{\odot}$ 
of H{\sc i}  per  beam; thus it would need to  have a column density, 
N$_{H{\sc i}}$, of at least 10$^{19}$  atoms cm$^{-2}$. In reality, our 
automatic baseline fitting procedure would at present almost certainly 
``remove'' the signal from the data. Irrespective of telescope size, 
 for sources that fill the beam  
there is a lowest column density 

\begin{equation}
 N_{\rm H{\sc i}} > 10^{18} \cdot T_{sys}{\sqrt{\frac{\Delta V [kms^{-1}]}{t_{obs} [s]}}}\hbox{cm}^{-2}
\end{equation}

\noindent  
detectable by any radio telescope, set by receiver noise, where 
$\Delta$V  is the projected linewidth in \kms (e.g. Disney \& Banks 
1997). 
Extreme objects, close to this column density limit, are best found when they 
are matched to the beam. For example, an  L$_{\star}$ galaxy ten times the 
radius of our own and containing the same amount of H{\sc i} would 
have an N$_{H{\sc i}} < 10^{19}$ atoms cm$^{-2}$ and would be 
found only in integrations significantly longer than \hipass, and 
most probably out near 6000\kms.  So the results from the present
H{\sc i} survey do not rule out the existence of extended LSB galaxies 
in the Cen\,A group.  The present study 
 is being followed by a much deeper H{\sc i} survey in  the same region, 
using the same instrument to address this issue. 

Although our survey has increased the number of known Cen\,A group 
members by 50\%, the new members add only 6\% to the total 
H{\sc i}, and 4\% to its light. The H{\sc i} mass function has 
a slope at the low mass end of  1.33 $\pm$ 0.15, which is unlikely to be
steeper than 1.5 due to either statistical or systematic effects.

 The five newly 
catalogued dwarfs have indicative mass to stellar mass ratios 
(col.[8], Table~3) between 40 and 80  (mean 63  $\pm$ 16). 
Such ratios probably underestimate the dynamical masses because they 
assume the H{\sc i} radii (R$_{H{\sc i}}$) are equal to the optical 
radii (R$_{opt}$). However, H{\sc i} maps of similar objects by other authors 
generally show that R$_{H{\sc i}} >$ R$_{opt}$  - the most 
striking example being the VLA observations of H{\sc i}1225+01, 
Chengalur, Giovanelli \& Haynes 1995  (for comparison, the five 
  new group members that are 
  previously catalogued 
  have a mean ratio of 14). The high ratios add to the 
accumulating evidence (e.g. Kormendy 1985) that H{\sc i}-rich 
dwarfs have their dynamics controlled by dark matter.

This survey of the Cen\,A group was aimed chiefly at validating 
the performance of the wider \hipass survey of the southern sky - 
which will be over 90\% complete by the time this goes to press. It 
confirms that performance matches expectation, and that large area 
blind H{\sc i} surveys will open a new window onto the extragalactic 
sky. As to the Cen\,A group itself, we expect to find further new group
members when we run the data through the automatic galaxy-finding 
algorithm now under development. We have also completed a much deeper 
survey (t$_{obs}$=12 $\times$ \hipass) of a limited region 
(8$^{\circ}$ x 4$^{\circ}$)  in the area, to  investigate the 
performance and promise of long multibeam integrations (Banks 1998). 

\acknowledgements

The development of the Parkes Multibeam Receiver was supported by 
the Australia Telescope National Facility, the Universities of 
Melbourne, Western Sydney, Sydney, Cardiff and  Manchester 
 and by  Mount 
Stromlo Observatory,  PPARC and the ARC. The receiver and hardware 
were constructed by the staffs of the CSIRO Division of Telecommunications 
and Industrial Physics with assistance from Jodrell Bank. The Multibeam 
software was developed by the AIPS++ team which included staff and students 
of the ATNF and University of Melbourne. 
The authors are very grateful to 
 an anonymous referee  whose comments  have led to 
 significant  changes in the 
 presentation of the results.  GDB acknowledges the 
 support of a PPARC postgraduate research award.

\newpage

\figcaption[cenplotmay.eps]{Known galaxies in the Cen\,A group. New Group members 
are indicated as filled circles, previously known group member are open circles. 
Our survey area lies inside the solid lines. C\^ot\'e's previous optical survey lay 
between $12^{h}$ $30^{m}<\alpha<15^{h}$ $00^{m}$ and $-20^{\circ}<\delta<-50^{\circ}$. \label{censpat}}

\figcaption{21-cm spectra of the 28 detected objects centered on the 
coordinates given in Table 1. Note that the ordinates differ between spectra. 
Asterisks as in table 1. The velocity range shown encompass the survey interval
which is between 200 and 1000 \kms .   The spectrum of CEN 5 is 
 included for completness but it is excluded from the statistics 
   because its velocity is below our 
 lower limit. 
\label{spec} }

\figcaption{Comparison of  integrated fluxes measured 
 by us and by  C\^ot\'e et al. (1997). If we exclude the 3 outlying 
 points - all higher for us because we 
 may pick up H{\sc i} not included in the C97 search area, the fluxes 
 agree within 10\%.  
  \label{flcomp}.}

\figcaption{Comparison of velocity widths between  C\^ot\'e et al. (1997) 
 and ourselves\label{velcomp}.}

\figcaption{optical B-band images of the 10 new Cen\,A group members, with 
asterisks as in table1. Orientation is  
East down, North to the left, except for 
 \hipass 1351-47 (East down, North right) and 
 ESO 174-G?001 (North down, East right) \label{optical}}

\figcaption{H{\sc i} mass function derived from the 
 detected   galaxies in the Cen\,A  group, omitting CEN 5, as being too close, and NGC 5128 as being an absorption measurement only.    
 Dashed lines indicate 1 standard deviation from fitted slope, passing through the weighted mean. \label{himf}}

\newpage

\tiny

\begin{deluxetable}{lllcrrrr}
\tablecaption{H{\sc i} Detections in the Centaurus A group \label{hiobs}}
\tablehead{
\colhead{Galaxy}&
\colhead{$\alpha_{j2000}$}&
\colhead{$\delta_{j2000}$}&
\colhead{$V_{\odot}$}&
\colhead{$\int{S \Delta V}$}&
\colhead{$\Delta$ V$_{50}$}&
\colhead{$\Delta$ V$_{20}$}&
\colhead{H{\sc i} Mass} \\
\colhead{}&
\colhead{(h m s)}&
\colhead{($^{\circ}$ $^{'}$ $^{''}$)}&
\colhead{(\kms)}&
\colhead{(Jy\,\kms)}&
\colhead{(\kms)}&
\colhead{(\kms)}&
\colhead{($\times$ 10$^{6}$ M$_{\odot}$)}\\
\colhead{(1)}&
\colhead{(2)}&
\colhead{(3)}&
\colhead{(4)}&
\colhead{(5)}&
\colhead{(6)}&
\colhead{(7)}&
\colhead{(8)}
}
\startdata
     ESO 321-G014*    & 12:13:56 & -38:14:05 & 610 &   6.4 &  35 & 43  & 19\\
     ESO 381-G018*    & 12:44:26 & -35:55:48 & 610 &   4.3 &  48 & 56  & 13\\
     ESO 381-G020    & 12:46:02 & -33:50:00 & 590 &  36 &  80 & 98  & 100\\
            CEN 5\tablenotemark{1}    & 13:03:25 & -49:32:48 & 130 &   5.9 &  21 & 46  &  17\\
            CEN 6    & 13:05:09 & -40:05:50 & 610 &   7.0 &  35 &  46 & 20\\
         NGC 4945    & 13:05:09 & -49:31:48 & 570 & 350 & 360 & 380 & 990\\
     ESO 269-G058*    & 13:10:33 & -46:59:52 & 400 &   7.2 &  62 & 84 & 21\\
   \hipass 1321-31**    & 13:21:06 & -31:32:25 & 570 &   8.4 &  31 & 46 & 24\\
         NGC 5102    & 13:21:55 & -36:38:18 & 470 & 110 & 200 & 220 & 320\\
      AM 1321-304    & 13:24:31 & -30:58:05 & 500 &   3.9 &  34 &  66  & 11\\
          IC 4247*    & 13:26:34 & -30:22:16 & 420 &   6.3 &  33 & 49  & 18\\
     ESO 324-G024    & 13:27:44 & -41:29:45 & 520 &  43 &  68 &  95 & 120\\
   \hipass 1328-30**    & 13:28:26 & -30:26:32 & 200 &  8.5 & 33 & 54   & 32\tablenotemark{2}\\
     ESO 270-G017    & 13:34:48 & -45:32:53 & 830 & 260 & 150 & 160 & 740\\
     NGC 5236 / M 83 & 13:37:03 & -29:54:37 & 510 & 1700 & 260 & 280 & 3700\\
   \hipass 1337-39**    & 13:37:26 & -39:52:15 & 490 &   7.9 &  37 & 51 & 23\\
         NGC 5237    & 13:37:40 & -42:50:29 & 360 &  12 &  75 &  92 & 33\\
         NGC 5253    & 13:39:58 & -31:39:03 & 400 &  51 &  65 & 97 & 150\\
         NGC 5264    & 13:41:34 & -29:54:19 & 500 &  28 &  43 & 98 &  82\\
     ESO 325-G011    & 13:45:03 & -41:51:43 & 550 &  27 &  58 & 81  & 77\\
    ESO 174-G?001*    & 13:48:01 & -53:20:46 & 680 &  73 &  72 & 100 & 210\\
   \hipass 1348-37**    & 13:48:47 & -37:58:28 & 570 &   4.3 &  30 & 40 & 12\\
     ESO 383-G087    & 13:49:20 & -36:02:59 & 330 &  31 &  33 & 54  & 90\\
   \hipass 1351-47**    & 13:51:12 & -46:58:13 & 530 &   4.4 &  38 & 47  & 13\\
         NGC 5408    & 14:03:21 & -41:23:17 & 500 &  55 & 75 & 120 & 
 160\\
    UKS1424-460    & 14:28:06 & -46:17:51 & 380 &  21 &  47 & 79  & 61\\
     ESO 222-G010    & 14:35:01 & -49:24:14 & 620 &   9.1 &  41 & 64  & 26\\
     ESO 274-G001    & 15:14:15 & -46:48:37 & 520 & 150 & 170 & 190 & 430\\
\multicolumn{6}{c}{} \\

NGC 5128\tablenotemark{3} &  13 25 28   &  -43 01 09           
  &  550    &  \nodata  & \nodata  &  \nodata  & 
 490.0\tablenotemark{4} \\  

 NGC 5206\tablenotemark{3}    &  13 33 44   &  -48 09 04           
  & 570 & \nodata & \nodata & \nodata &  $<$ 5.5\tablenotemark{5}  \\ 

ESO 272-G025\tablenotemark{3} &  14 43 26  &  -44 42 19 &  
 620  &  \nodata 
 &  \nodata  & \nodata  &  $<$ 5.5\tablenotemark{5}  \nl

\enddata

\tablenotetext{}{* Previously catalogued but not known to be group members.}

\tablenotetext{}{** Previously not catalogued.}

\tablenotetext{1} {Not included in the statistics because v$_{\odot} <$ 
 200 \kms.}

\tablenotetext{2}  {H{\sc i} mass and linewidths uncertain, galaxy profile confused 
with HVC.} 

\tablenotetext{3}{Undetected in \hipass, as below our limiting sensitivity.  
$\alpha$, $\delta$ and heliocentric radial velocity obtained  from NED.}

\tablenotetext{4}{mass limit from RC3 (de vaucoulers et al.  1991). }

\tablenotetext{5}{H{\sc i} limit from  C\^ot\'e et al.  1997, who 
 detected H$\alpha$.}

 \end{deluxetable}

\normalsize

\begin{deluxetable}{llllllll}

\scriptsize
\tablewidth{0pt}
\tablecaption{Photometric Parameters of the new Centaurus\,A Galaxies \label{photomone} }
\tablehead{ 
\colhead{Galaxy}  & 
\colhead{$\alpha_{j2000}$ } & 
\colhead{ $\delta_{j2000}$} & 
\colhead{m$_{B}$} & 
\colhead{D$(B_{25})$ }&  
\colhead{$\mu_{0}^{exp}$} & 
\colhead{ $r_0$ }&
\colhead{ $r_0$}  \\  
\colhead{ } &
\colhead{(h m s) } &
\colhead{ ($^{\circ}$ $^{'}$ $^{''}$)} &   
\colhead{(mag)}&  
\colhead{($''$) } & 
\colhead{(mag arcsec$^{-2}$)} & 
\colhead{ ($''$)} & 
\colhead{ (kpc)} \\ 
\colhead{(1)} &
\colhead{(2) } &
\colhead{ (3)} &   
\colhead{(4)}& 
\colhead{(5)}& 
\colhead{(6)}&  
\colhead{(7) } & 
\colhead{(8)}\\ 
}
\startdata 
ESO321-G014 & 12:13:50 & -38:13:53 & 15.2\tablenotemark{1} & \nodata 
&  \nodata & \nodata   & \nodata  \nl
ESO381-G018 & 12:44:42 & -35:58:07 & 15.8\tablenotemark{1}  &  22 &  22.8 & 10  &  0.2  \nl
ESO269-G058 & 13:10:33 & -46:59:22  & 12.2$\pm$0.2 &  108 & 21.4 & 30 & 0.5  \nl
\hipass 1321-31   & 13:21:08  & -31:31:48 & 17.1$\pm$0.2 & 8 & 24.2 &  10 &  0.2  \nl
IC 4247     & 13:26:44 & -30:21:45 & 14.4\tablenotemark{1} & \nodata & 
 \nodata & \nodata & \nodata \nl
\hipass 1328-30   & 13:28:55 &  -30:29:32 & 17.0$\pm$0.2 &  16 &  23.4 &  10 & 0.2   \nl
\hipass 1337-39   & 13:37:26 &  -39:53:47 & 16.5$\pm$0.3\tablenotemark{2} & \nodata &  
\nodata &  \nodata &  \nodata \nl
ESO 174-G?001\tablenotemark{3} & 13:47:50 & -53:20:45 & 14.2 & \nodata &  \nodata & 
\nodata & \nodata  \nl
\hipass 1348-37  &  13:48:33 & -37:58:03 & 16.9$\pm$0.2  &   11 &  23.8 &  9 & 0.2  \nl
\hipass 1351-47  \tablenotemark{3}   & 13:51:22 & -47:00:07 &  17.5$\pm$0.3  & \nodata & 
 \nodata & \nodata & \nodata  \nl
\enddata

\tablenotetext{1}{m$_{B}$ values from RC3 (de Vaucouleurs, 1991)} 

\tablenotetext{2} {Calibrated with B-band data obtained from DSS}

\tablenotetext{3} {Object is too faint (or low in the Galactic plane) to 
 be fit by a light profile.}

\end{deluxetable}

\normalsize

\begin{deluxetable}{lllllllll}
\tablewidth{0pt}
\tablecaption{Intrinsic Parameters of the new Centaurus\,A Galaxies\label{photomtwo}}
\tablehead{ 
\colhead{Galaxy}   &  
\colhead{$R_{opt}$} &
\colhead{ m$_{T}$} &
\colhead{A$_{B}$} &  
\colhead{M$_{B}$} &  
\colhead{inc.} &   
\colhead{M$_{\rm H{\sc i}}$/L$_{\rm B}$} &  
\colhead{M$_{\star}$} &   
\colhead{M$_{\rm i}$/M$_{\star}$} \\  
\colhead{} & 
\colhead{(arcmin )} &  
\colhead{(mag)} &  
\colhead{(mag)} & 
\colhead{(mag)}&
\colhead{(deg.)}&     
\colhead{} &  
\colhead{ (10$^{8}$ M$_{\odot}$)} &   
\colhead{}\\ 
\colhead{(1)} & 
\colhead{(2)} &  
\colhead{(3)} &  
\colhead{(4)} & 
\colhead{(5)}&     
\colhead{(6)} &  
\colhead{(7)} &
\colhead{(8)} &   
\colhead{(9)} 
}
\startdata
ESO 321-G014         &1.4$\pm$0.2 &\nodata & 0.40 &-13.0 & 62 & 0.83 & 0.22 & 22$^{+ 15}_{-10}$  \nl
ESO381-G018          &1.1$\pm$0.2 & 15.8   & 0.27 &-12.2 & 55 &  1.1  & 0.11 &  77$^{+40}_{-29}$  \nl
ESO 269-G058         &3.8$\pm$0.3 & 12.0   & 0.48 &-15.1 & 53 &  0.13 & 1.6  &  32$^{+14}_{-11}$  \nl
\hipass 1321-31 &0.7$\pm$0.2 & 17.2   & 0.27 &-10.9 & 52 &  7.4  & 0.033& 74$^{+56}_{-37}$  \nl
IC 4247              &1.2$\pm$0.1 &\nodata & 0.24 &-13.6 & 63 &  0.44 & 0.40 &  9$^{+5}_{-4}$    \nl
\hipass 1328-30 &0.9$\pm$0.1 & 16.4   & 0.32 &-11.0 & 57 &  8.4  & 0.038& 84$^{+48}_{-35}$  \nl
\hipass 1337-39 &0.5$\pm$0.2 &\nodata & 0.37 &-11.5 & 42 &  3.8  & 0.060& 61$^{+56}_{-35}$ \nl
ESO 174-G?001        &2.7$\pm$0.3 &\nodata & 2.11 &-15.6 & 76 &  0.80 &  2.6 &   13$^{+5}_{-4}$  \nl
\hipass 1348-37 &0.5$\pm$0.2 & 17.0   & 0.33 &-11.2 & 64 &  2.7  & 0.042&  37$^{+35}_{-22}$ \nl
\hipass 1351-47 & 0.6$\pm$0.2 & \nodata  & 0.65 &-10.9 & \nodata &  3.9 & 0.033 &   
59$^{+44}_{-30}$ \nl
\enddata
\end{deluxetable}

\begin{deluxetable}{llrrl}
\tablecaption{Comparison of Faint End Mass Functions \label{othersurvs}}
\tablehead{
\colhead{Survey}&
\colhead{N}&
\colhead{No. $<$ 10$^{8}$ M$_{\odot}$}&
\colhead{M$^{\star}_{H{\sc i}}$/M$_{\odot}$}&
\colhead{$\alpha$}\\   
\colhead{(1)}&
\colhead{(2)}&
\colhead{(3)}&
\colhead{(4)}&
\colhead{(5)}   
}
\startdata
 This work   & 27  & 17 &  \nodata   & 1.30 $\pm$ 0.15 \\
Spitzak \& Schneider (1998)       & 79  &  4 &  7 $\times$ 10$^{9}$   & 1.32  \\
Zwaan et al. (1997)  & 66  &  4 &  3.5 $\times$ 10$^{9}$ & 1.20 \\
Kilborn et al. (1998) & 99 &  1 &  3 $\times$ 10$^{9}$   & 1.35 \\
Kilborn et al. (1999)\tablenotemark{1} & 263 & 3 & 3 $\times$ 10$^{9}$ & 1.20 \\
Kilborn et al. (1999)\tablenotemark{2} & 253 & 3 & (5.5$^{+2.2}_{-1.1}$) $\times$ 10$^{9}$ & 1.15 $\pm$ 0.17 \\
\enddata
\tablenotetext{1}{$\Sigma\,1/V_{max}$ method}
\tablenotetext{2}{Maxiumum likelihood method}
\end{deluxetable}


\newpage

\begin{references}

\reference {} Banks, G.D. 1998, ``21-cm Blind Searches for Galaxies'', PhD thesis, University of Wales  

\reference {} Barnes, D.G., Staveley-Smith, L., Ye, T. \& Oosterloo, 
T. 1998, ADASS VII (San Francisco: ASP)

\reference {} Bergvall, N. \& J\"ors\"ater, S.  1988, Nature, 331, 589


\reference{} Bevington, P.R. 1969, ``Data Reduction and Error Analysis 
 for the Physical Science'', McGraw-Hill, p107-116


\reference{} Bosma, A. 1978, PhD Thesis, University of Groningen

\reference {} Brinks E.  \& Klein U.  1988, MNRAS, 231, 63

\reference {a8} C\^ot\'e, S., Freeman K.C., Carnigan, C. \& Quinn, P.
1997, AJ, 114, 1313

\reference{} Chengalur J.N., Giovanelli, R. \& Haynes, M.P. 1995, AJ, 109, 1729


\reference{}  da Costa, L.N.,   Willmer, C.,  Pellegrini, P.S. \& Chincarini, G. 1987, AJ, 93, 1338



\reference{}  de Blok, W.J.G., McGaugh, S.S. \& 
 van der Hulst, J.M. 1996, MNRAS, 283, 18


\reference{}  de Propris, R., Pritchet, C.J., Harris, W.E. \& 
  McClure, R.D.  1995, AJ, 450, 534

\reference {} de Vaucouleurs, G. 1979, AJ, 84, 1270

\reference {} de Vaucouleurs, G., de Vaucouleurs, A., Corwin Jr., H.G., 
Buta, R.J., Paturel, G. \& Fouqu\'{e}, P. 1991, Third Reference Catalogue of 
Bright Galaxies



\reference{} Dickey, J.M. 1997, AJ, 113, 1939

\reference{}  Disney, M.J 1998, IAU Symposium 179, ``New Horizons from 
Multi-Wavelength Sky Surveys'',   ed.  B.J. McLean et al. , 
Kluwer Academic Publishers, p11



\reference{}  Disney, M.J. \& Banks, G.D. 1997, Proc. A. S. Australia, 69, 14

\reference{} Draper, P. 1993, CCD Data Reduction Pack - Starlink 
User Note 139.3, Starlink-Rutherford Appleton Labs. Didcot, UK

\reference {} Driver, S. P. \& Phillipps S.  1996, ApJ, 469, 529

\reference {} Freeman K.C. 1997,  Proc. A. S. Australia, 14, 4

\reference {} Graham J.A.  1979, ApJ, 232, 60

\reference {} Haynes M.P.  \&  Roberts M.S. 1979, ApJ, 227, 767



\reference{}  Henning, P.A. 1995, AJ,   450, 578 


 \reference{}  Huchtmeier, W.K., Karachentsev, I.D. \& 
 Karachentseva, V.E. 1997, A\&A, 322, 375


\reference{} Huchtmeier, W.K. \& Skillman E.D. 1998, A \& AS, 127, 397

\reference {} Hui, X., Ford, H.C., Ciardullo, R.  \& Jacoby, G. H. 
 1993,  ApJ,   414, 463 


\reference{} Impey, C.D. \& Bothun, G.D. 1997, ARA\&A,  35, 267 



\reference{} Karachentseva, V.E.  \& Karachentsev, I.D.,  1998, 
 A \& AS, 127, 409

\reference{}  Kerr, F.J. \& Henning, P.A. 1987, ApJlet, 320, 99

\reference{} Kilborn, V., Staveley-Smith, L. 
 \& Webster, R.  1998, Proceedings of the XVIIIth 
Rencontre de Moriond `Dwarf galaxies and Cosmology', eds. T.X.Thuan, 
C. Balkowski, V.Cayatte, J.Tran Than Van, 1998 Editions Frontieres, in press

\reference{} Kilborn, V., Webster, R. \& Staveley-Smith, L. 1999, PASA, 16, in press

\reference{} Kormendy, J.  1985, ApJ, 295, 73

\reference{}  Krumm,N. \& Brosch, N.  1984, AJ, 89, 1461

\reference{} Landolt, A.\ U.\ 1992, \aj, 104, 340

\reference{} Lauberts, A.  1982, The ESO/UPPSALA Survey  of 
the ESO(B) Atlas, Munich:ESO

\reference{} Lequeux, J. \& Viallefond, F.  1980, A\&A, 91, 269

\reference{} Lin, H.,  Kirshner, R.P.,  Schectman, S.A., Landy, S.D.,   
Oemler,A., Tucker, D.L.   \& Schechter, P.L. 1996, ApJ, 464, 60

\reference {}  Lo, K. Y.  \& Sargent, W.L.W. 1979, ApJ, 227, 756 

\reference{}  Loveday, J.,  Peterson, B.A., Efstathiou, G.,  \& Maddox, S.J.  
 1992, ApJ, 390, 338

\reference{}   Maia, M.A.G. \& Willmer, C.N.A. 1997, in IAU Symposium 
 no. 186 ``Galaxy Interactions at Low and High redshift'', 
 Kyoto, Japan 

\reference{} Matthews, L. \& Gallagher, J.  1996, AJ, 111, 1098



\reference{} McGaugh, S.S. 1996,  MNRAS, 280, 337


\reference{}  McMahon, P., van Gorkom, J., Richter, O. \& Ferguson, H.  
    1992, in  ``The Evolution of Galaxies and their Environment'', 
    NASA Conference Publications 3190, eds. 
  Hollenbach, D., et al.



\reference{} Oosterloo, T. \& Iovino, A. 1997, PASA, 14, 89

 
\reference {} Phillipps, S., Parker, Q.A., Schwartzenberg, J.M.   
\& Jones, J.B. 1998, ApJlet, 493, 59


\reference{} Roberts, M.S.   1978, AJ, 83, 1026


\reference{} Saha, A., Sandage, A., Labhardt, L., 
 Schwengeler, H., Tammann, G.A., Panagia, N. 
 \&  Macchetto, F.D.  1995,  ApJ, 438, 8 


\reference {} Sandage, A., Saha, A., Tammann, G. A., 
Labhardt, L., Schwengeler, H., Panagia, N., \& Macchetto, F. D. 
1994, ApJ, 425, 14

\reference{} Sault, R.J.,  Tenben, P.J. \& Wright, M.C.H. 
1995,  in ``Astronomical Data Analysis Software and Systems IV'', 
  ASP Conference Series, 77, ed. R.A.  Shaw,  H.E. 
Payne and J.J.E. Haynes,   ASP. San Francisco, p 33

\reference {} Schechter, P. 1976, ApJ, 203, 297


\reference{} Schlegel D.J., Finkbeiner, D.P., \& Davis, M. 1998, ApJ, 500, 525 




\reference{} Spitzak, J.G. \& Schneider, S.E. 1998, astro-ph/9811336



\reference{}  Staveley-Smith, L.,  Bland, J.,  Axon, D.J.,  Davies, R.D. \&  Sharples, R.M. 1990, AJ, 364, 23



\reference {a2} Staveley-Smith, L., Wilson W.E., Bird T.S., 
 Disney M.J., Ekers R.D., Freeman K.C., Haynes R.F., 
Sinclair M.W., Vaile R. A., Webster R. L. \&  Wright A.E.     
1996, Proc. A. S. Australia, 13, 243

\reference {} Staveley-Smith, L., 
 Webster, R.L., Banks, G.D., 
 Kilborn, V., Koribalski, B. \&  Putman, M.   
1999, ``The Local Galaxy Population from the 
\hipass Survey'',  
  in ESO/ATNF International Workshop ``Looking 
Deep in the Southern Sky'', eds. R. Morganti \& W. Couch, 
 248


 \reference{} Trentham, N.  1997, MNRAS, 286, 133

\reference{} Turner, J.A., Phillipps, S., Davies, J.I. \&  
  Disney, M.J.    1993, MNRAS, 261, 39

\reference{} Wakker B. 1991, A\&A, 250, 509

\reference{} Webster, B.L., Goss, W.M., Hawarden, T.G., Longmore A.J.  
\& Mebold, U. 1979, MNRAS, 186, 31


\reference{}  Weinberg, D.H.,  Szomoru, A.,  Guhathakurta, P. 
 \&   van Gorkom, J.H. 1991, ApJlet, 372, 13

\reference{}  van Driel, W., Kraan-Korteweg, R.C., Bingelli, B. 
 \& Huchtmeier, W.K. 1998 A\&AS, 127, 397   

\reference{} van Gorkom J.H., van der Hulst, J.M., Haschick, 
 A.D., \& Tubbs, A.D. 1990, 
 AJ, 99, 1781

\reference{} van Zee L., Haynes M.P.  \& Giovanelli R. 1995, AJ, 109, 990


\reference{}  Williams, B.A., McMahon, P.M. \& 
 van Gorkom, J.H. 1991, AJ, 101, 1957 

\reference{}   Williams, B.A.  \& van Gorkom, J.H. 1997, in  
 IAU Symposium 
 no. 186 ``Galaxy Interactions at Low and High redshift'', 
 Kyoto, Japan  
 
 

\reference {} Zwaan, M., Briggs, F.H.,  Sprayberry, D.   \& Sorar, E. 
1997, ApJ, 490, 173

\end{references}
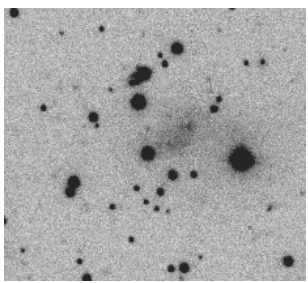
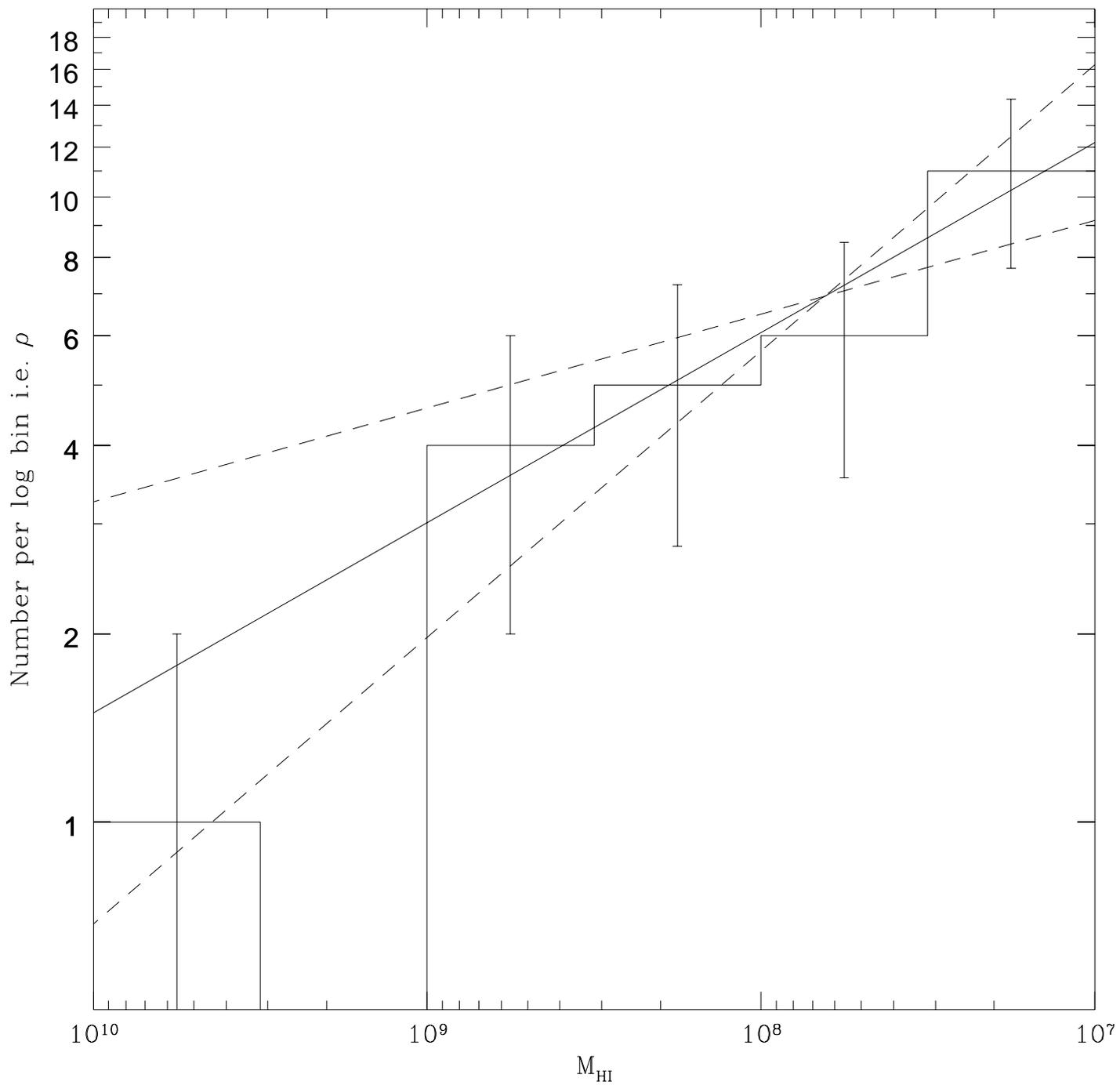
\end{document}